\documentstyle[sprocl]{article}
\def\be{\begin{equation}}
\def\ee{\end{equation}}
\def\bea{\begin{eqnarray}}
\def\eea{\end{eqnarray}}

\begin{document}
\title{NEUTRINO MASSES AND ANOMALOUS INTERACTIONS}
\author{ERNEST MA}
\address{Department of Physics, University of California\\
Riverside, CA 92521, USA}
\maketitle\abstracts{The interplay between neutrino masses and the 
interactions of neutrinos with matter is discussed with an eye to 
extending the latter to include possible new interactions.  This 
conjecture may resolve the conundrum posed by the present experimental 
data on neutrino oscillations which suggest the existence of 
four neutrinos, whereas $Z$ decay and Big Bang Nucleosynthesis allow 
only three.  The case of a possible sterile neutrino is also briefly 
discussed.}

\section{Neutrino Masses}

In the minimal standard model, under the gauge group $SU(3)_C \times SU(2)_L 
\times U(1)_Y$, the quarks and leptons transform as:
\begin{equation}
\left[ \begin{array} {c} u \\ d \end{array} \right]_L \sim (3, 2, 1/6), ~~~ u_R 
\sim (3, 1, 2/3), ~~~ d_R \sim (3, 1, -1/3);
\end{equation}
\begin{equation}
\left[ \begin{array} {c} \nu \\ l \end{array} \right]_L \sim (1, 2, -1/2), 
~~~ l_R \sim (1, 1, -1).
\end{equation}
There is also the Higgs scalar doublet $(\phi^+, \phi^0) \sim (1, 2, 1/2)$ 
whose nonzero vacuum expectation value $\langle \phi^0 \rangle = v$ breaks 
$SU(2)_L \times U(1)_Y$ to $U(1)_Q$.  Note that the existence of the term
\begin{equation}
\overline {(\nu, l)}_L l_R \left[ \begin{array} {c} \phi^+ \\ \phi^0 
\end{array} \right] ~~ \Longrightarrow ~~ m_l \neq 0.
\end{equation}
However, the absence of $\nu_R$ implies that $m_\nu = 0$.

The canonical way of obtaining small neutrino masses is via the seesaw 
mechanism~\cite{1}.  This assumes the existence of one $\nu_R \sim (1, 1, 0)$ for 
each $\nu_L$, so that we have the terms
\begin{equation}
{m_D \over v} \overline {(\nu, l)}_L \nu_R \left[ \begin{array} {c} 
\bar \phi^0 \\ -\phi^- \end{array} \right] ~~ {\rm and} ~~ {1 \over 2} 
M \nu_R \nu_R.
\end{equation}
Thus the $2 \times 2$ neutrino mass matrix linking $\bar \nu_L$ to $\nu_R$ 
is given by
\begin{equation}
{\cal M} = \left[ \begin{array} {c@{\quad}c} 0 & m_D \\ m_D & M \end{array} 
\right] ~~ \Longrightarrow ~~ m_\nu = {m_D^2 \over M}.
\end{equation}
Here, $\nu_L - \nu_R^c$ mixing is $m_D/M$ and $M$ is the scale of new 
physics.  In this minimal scenario, new physics enters only through $M$, 
hence there is no other observable effect except for a nonzero $m_\nu$. 
Actually, $m_D/M$ is in principle observable but it is in practice far too 
small.

\section{Neutrino Oscillations}

It is well-known that neutrinos may oscillate into one another if their mass 
eigenstates do not coincide with their interaction eignestates.  Let
\begin{eqnarray}
\nu_e &=& \nu_1 \cos \theta + \nu_2 \sin \theta, \\ \nu_\mu &=& -\nu_1 \sin 
\theta + \nu_2 \cos \theta,
\end{eqnarray}
where $m_{\nu_1} \neq m_{\nu_2}$, then if $\nu_e$ is created at $t=0$, 
$\vec x = 0$; at a later $t$ and $\vec x$ away, $\nu_e$ becomes
\begin{equation}
e^{-i(E_1 t - \vec p \cdot \vec x)} \nu_1 \cos \theta + e^{-i(E_2 t - \vec p 
\cdot \vec x)} \nu_2 \sin \theta.
\end{equation}
The probability that this state would be measured as $\nu_e$ (by producing 
$e$) is
\begin{equation}
|\cos^2 \theta + e^{-i(E_2-E_1)t} \sin^2 \theta|^2 = 1 - {1 \over 2} \sin^2 
2 \theta [1 - \cos (E_2-E_1)t ],
\end{equation}
where $E_{1,2} \simeq p (1 + m_{1,2}^2/2p^2)$.  Now
\begin{equation}
(E_2 - E_1) t \simeq {{(m_2^2 - m_1^2)t} \over {2p}} \simeq {{(\Delta m^2) L} 
\over {2E}},
\end{equation}
hence the probability that $\nu_e$ remains $\nu_e$ is given by
\begin{equation}
P(\nu_e \rightarrow \nu_e) = 1 - {1 \over 2} \sin^2 2 \theta \left[ 1 - \cos 
{{(\Delta m^2) L} \over {2E}} \right].
\end{equation}
Note that the sign of $\Delta m^2$ does not matter in this case.

In traversing matter, neutrinos interact with the electrons and nuclei of 
the intervening material and their forward coherent scattering induces an 
effective mass analogous to the occurrence of an index of refraction for light, 
and may result in the resonance conversion of one flavor to another, {\it 
i.e.} the famous MSW effect~\cite{2}.  There are two kinds of known 
interactions: (A) the exchange of a charged $W$ boson between $\nu_e$ and 
$e$, and (B) the exchange of a neutral $Z$ boson between $\nu_e$ or $\nu_\mu$ 
with electrons and quarks.  The latter is identical for all neutrino flavors. 
The evolution equation for neutrino oscillations in matter is then given by
\begin{equation}
-i {d \over {dt}} |\nu \rangle_{e,\mu} = \left( p + {{\cal M}^2 \over {2p}} 
\right) |\nu \rangle_{e,\mu},
\end{equation}
where
\begin{equation}
{\cal M}^2 = {\cal U} \left[ \begin{array} {c@{\quad}c} m_1^2 & 0 \\ 0 & 
m_2^2 \end{array} \right] {\cal U}^\dagger + \left[ \begin{array} {c@{\quad}c} 
A + B & 0 \\ 0 & B \end{array} \right].
\end{equation}
In the above, the charged-current interaction $A$ applies only to $\nu_e$ 
and is given by
\begin{equation}
A = \pm 2 \sqrt 2 G_F N_e p \left\{ \begin{array} {c} + ~{\rm for}~ \nu \\ 
- ~{\rm for}~ \bar \nu \end{array} \right\},
\end{equation}
where $N_e$ is the number density of electrons in matter.  For
\begin{equation}
{\cal U} = \left( \begin{array} {c@{\quad}c} c & s \\ -s & c \end{array} 
\right),
\end{equation}
we get
\begin{equation}
{\cal M}^2 = \left[ \begin{array} {c@{\quad}c} c^2 m_1^2 + s^2 m_2^2 + A + B 
& sc(m_2^2 - m_1^2) \\ sc (m_2^2 - m_1^2) & s^2 m_1^2 + c^2 m_2^2 + B 
\end{array} \right].
\end{equation}
A resonance occurs when ${\cal M}^2_{11} = {\cal M}^2_{22}$, namely
\begin{equation}
(m_2^2 - m_1^2) \cos 2 \theta - A = 0.
\end{equation}
Since $A > 0$ for $\nu_e$ coming from the sun, $m_2 > m_1$ is required 
for the MSW effect.

\section{Three Neutrinos versus Four Neutrinos}

Present experimental evidence for neutrino oscillations~\cite{3} includes 
the solar $\nu_e$ deficit which requires $\Delta m^2$ of around $10^{-5}$ 
eV$^2$ for the MSW explanation or $10^{-10}$ eV$^2$ for the 
vacuum-oscillation solution, the atmospheric neutrino deficit in the 
ratio $\nu_\mu + \bar \nu_\mu / \nu_e + \bar \nu_e$ which implies a 
$\Delta m^2$ of around $10^{-2}$ eV$^2$, and the LSND experiment which 
indicates a $\Delta m^2$ of around 1 eV$^2$.  Three different $\Delta m^2$ 
necessitate four neutrinos, but the invisible width of the $Z$ boson as well 
as Big Bang Nucleosynthesis allow only three.  If all of the above-mentioned 
experiments are interpreted correctly as due to neutrino oscillations, we 
are faced with a theoretical challenge in trying to understand how 
three can equal four.

One possibility is that there is a light singlet neutrino in addition to 
the three known doublet neutrinos $\nu_e$, $\nu_\mu$, and $\nu_\tau$. 
If so, why is this singlet light and how does it mix with the other three 
neutrinos?  Both questions can be answered in a model~\cite{4} based on 
$E_6$ inspired by superstring theory.  In the fundamental {\bf 27} 
representation of $E_6$, outside the 15 fields belonging to the minimal 
standard model, there are 2 neutral singlets.  One ($N$) is identifiable 
with the right-handed neutrino because it is a member of the {\bf 16} 
representation of $SO(10)$; the other ($S$) is a singlet also under $SO(10)$. 
In the reduction of $E_6$ to $SU(3)_C \times SU(2)_L \times U(1)_Y$, an extra 
U(1) gauge factor may survive down to the TeV energy scale.  It could be chosen 
such that $N$ is trivial under it, but $S$ is not.  This means that $N$ 
is allowed to have a large Majorana mass so that the usual seesaw mechanism 
works for the three doublet neutrinos.  At the same time, $S$ is protected 
from having a mass by the extra U(1) gauge symmetry, which I call $U(1)_N$. 
However, it does acquire a small mass from an analog of the usual seesaw 
mechanism because it can couple to doublet neutral fermions which are 
present in the {\bf 27} of $E_6$ outside the {\bf 16} of $SO(10)$.  
Furthermore, the mixing of $S$ with the doublet neutrinos is also possible 
through these extra doublet neutral fermions.  For details, see Ref.~[4].

\section{Three Neutrinos and One Anomalous Interaction}

If one insists on keeping only the usual three neutrinos and yet try to 
accommodate all present data, how far can one go?  It has been pointed out 
by many authors~\cite{5} that both solar and LSND data can be explained, 
as well as most of the atmospheric data except for the zenith-angle 
dependence.  It is thus worthwhile to consider the following scenario~\cite{6} 
whereby a possible anomalously large $\nu_\tau$-quark interaction may mimic 
the observed zenith-angle dependence of the atmospheric data.  Consider first 
the following approximate mass eigenstates:
\begin{eqnarray}
\nu_1 &\sim& \nu_e ~~~ {\rm with} ~ m_1 \sim 0, \\ \nu_2 &\sim& c_0 \nu_\mu 
+ s_0 \nu_\tau ~~~ {\rm with} ~ m_2 \sim 10^{-2} ~{\rm eV}, \\ \nu_3 &\sim& 
-s_0 \nu_\mu + c_0 \nu_\tau ~~~ {\rm with} ~ m_3 \sim 0.5 ~{\rm eV},
\end{eqnarray}
where $c_0 \equiv \cos \theta_0$, $s_0 \equiv \sin \theta_0$, and $\theta_0$ 
is not small.  Allow $\nu_1$ to mix with $\nu_3$ with a small angle 
$\theta '$ and the new $\nu_1$ to mix with $\nu_2$ with a small angle 
$\theta$, then the LSND data can be explained with $\Delta m^2 \sim 0.25$ 
eV$^2$ and $2 s_0 s' c' \sim 0.19$ and the solar data can be understood as 
follows.

Consider the basis $\nu_e$ and $\nu_\alpha \equiv c_0 \nu_\mu + s_0 \nu_\tau$. 
Then the analog of Eq.~(12) holds with Eq.~(13) replaced by
\begin{equation}
{\cal M}^2 = {\cal U} \left[ \begin{array} {c@{\quad}c} 0 & 0 \\ 0 & m_2^2 
\end{array} \right] {\cal U}^\dagger + \left[ \begin{array} {c@{\quad}c} 
A+B & 0 \\ 0 & B+C \end{array} \right],
\end{equation}
where $C$ comes from the anomalous $\nu_\tau$-quark interactions in the 
sun. 
The resonance condition is now
\begin{equation}
m_2^2 \cos 2 \theta - A + C = 0,
\end{equation}
where~\cite{7}
\begin{equation}
A - C = 2 \sqrt 2 G_F (N_e - s_0^2 \epsilon'_q N_q) p.
\end{equation}
In order to have a large $\epsilon'_q$ and yet satisfy the resonance condition 
for solar-neutrino flavor conversion, $m_2$ should be larger than its 
canonical value of $2.5 \times 10^{-3}$ eV, and $\epsilon'_q$ should be 
negative. [If $\epsilon'_q$ comes from $R$-parity violating squark exchange, 
then it must be positive, in which case an inverted mass hierarchy, {\it i.e.} 
$m_2 < m_1$ would be needed.  If it comes from vector exchange, it may be of 
either sign.]  Assuming as a crude approximation that $N_q \simeq 4 N_e$ in 
the sun, the usual MSW solution with $\Delta m^2 = 6 \times 10^{-6}$ eV$^2$ 
is reproduced here with
\begin{equation}
s_0^2 \epsilon'_q \simeq -3.92 = -4.17 (m_2^2/10^{-4}{\rm eV}^2) + 0.25.
\end{equation}
The seemingly arbitrary choice of $\Delta m_{21}^2 \sim 10^{-4}$ eV$^2$ is 
now sen as a reasonable value so that $\epsilon'_q$ can be large enough to 
be relevant for the following discussion on the atmospheric neutrino data.

Atmospheric neutrino oscillations occur between $\nu_\mu$ and $\nu_\tau$ 
in this model with $\Delta m^2_{32} \sim 0.25$ eV$^2$, the same as for 
the LSND data, but now it is large relative to the $E/L$ ratio of the 
experiment, hence the cosine factor of Eq.~(11) washes out and
\begin{equation}
P_0 (\nu_\mu \rightarrow \nu_\mu) = 1 - {1 \over 2} \sin^2 2 \theta_0 
\simeq 0.66 ~~ {\rm for} ~~ s_0 \simeq 0.47.
\end{equation}
In the standard model, this would hold for all zenith angles.  Hence it 
cannot explain the present experimental evidence that the depletion is more 
severe for neutrinos coming upward to the detector through the earth than for 
neutrinos coming downward through only the atmosphere.  This zenith-angle 
dependence appears also mostly in the multi-GeV data and not in the sub-GeV 
data.  It is this trend which determines $\Delta m^2$ to be around 
$10^{-2}$ eV$^2$ in this case.  As shown below, the hypothesis that $\nu_\tau$ 
has anomalously large interactions with quarks will mimic this zenith-angle 
dependence even though $\Delta m^2$ is chosen to be much larger, {\it i.e.} 
0.25 eV$^2$.

Consider the basis $\nu_\mu$ and $\nu_\tau$.  Then the analog of Eq.~(12) 
holds with Eq.~(13) replaced by
\begin{equation}
{\cal M}^2 = {\cal U}_0 \left[ \begin{array} {c@{\quad}c} 0 & 0 \\ 0 & m_3^2 
\end{array} \right] {\cal U}_0^\dagger + \left[ \begin{array} {c@{\quad}c} 
B & 0 \\ 0 & B + C \end{array} \right].
\end{equation}
The resonance condition is then
\begin{equation}
m_3^2 \cos 2 \theta_0 + C = 0,
\end{equation}
where $N_q$ in $C$ now refers to the quark number density inside the earth 
and the factor $s_0^2$ in Eq.~(23) is not there.  If $C$ is large enough, 
the probability $P_0$ would not be the same as the one in matter.  Using 
the estimate $N_q \sim 9 \times 10^{30}$ m$^{-3}$ and defining
\begin{equation}
X \equiv \epsilon'_q E_\nu/(10 ~{\rm GeV}),
\end{equation}
the effective mixing angles in matter are given by
\begin{eqnarray}
\tan 2 \theta_m^E &=& {{\sin 2 \theta_0} \over {\cos 2 \theta_0 + 0.091 X}} 
~~ {\rm for} ~ \nu, \\ \tan 2 \bar \theta_m^E &=& {{\sin 2 \theta_0} \over 
{\cos 2 \theta_0 - 0.091 X}} ~~ {\rm for} ~ \bar \nu.
\end{eqnarray}
For sub-GeV neutrinos, $X$ is small so matter effects are not very important. 
For multi-GeV neutrinos, $X$ may be large enough to satisfy the resonance 
condition of Eq.~(27).  Assuming adiabaticity, the neutrino and antineutrino 
survival probabilities are given by
\begin{eqnarray}
P(\nu_\mu \rightarrow \nu_\mu) &=& {1 \over 2} (1 + \cos 2 \theta_0 \cos 2 
\theta_m^E), \\ \bar P(\bar \nu_\mu \rightarrow \bar \nu_\mu) &=& {1 \over 2} 
(1 + \cos 2 \theta_0 \cos 2 \bar \theta_m^E).
\end{eqnarray}
Since $\sigma_\nu \simeq 3 \sigma_{\bar \nu}$, the observed ratio of 
$\nu + \bar \nu$ events is then
\begin{equation}
P_m \simeq {{3 r P + \bar P} \over {3 r + 1}},
\end{equation}
where $r$ is the ratio of the $\nu_\mu$ to $\bar \nu_\mu$ flux in the upper 
atmosphere.  The atmospheric data are then interpreted as follows.  For 
neutrinos coming down through only the atmosphere, $P_0 = 0.66$ applies. 
For neutrinos coming up through the earth, $P_m \simeq P_0 \simeq 
0.66$ as well for the sub-GeV data.  However, for the multi-GeV data, 
if $X = -15$, then $P = 0.31$ and $\bar P = 0.76$, hence $P_m$ is lowered 
to 0.39 if $r = 1.5$ or 0.42 if $r = 1.0$.  The apparent zenith-angle 
dependence of the data may be explained.

\section{Conclusion and Outlook}

If all present experimental indications of neutrino oscillations turn out to 
be correct, then either there is at least one sterile neutrino beyond the 
usual $\nu_e$, $\nu_\mu$, and $\nu_\tau$, or there is an anomalously large 
$\nu_\tau$-quark interaction.  The latter can be tested at the forthcoming 
Sudbury Neutrino Observatory (SNO) which has the capability of neutral-current 
detection.  The predicted $\Delta m^2$ of 0.25 eV$^2$ in $\nu_\mu$ to $\nu_e$ 
and $\nu_\tau$ oscillations will also be tested at the long-baseline neutrino 
experiments such as Fermilab to Soudan 2 (MINOS), KEK to Super-Kamiokande 
(K2K), and CERN to Gran Sasso.

More immediately, the new data from Super-Kamiokande, Soudan 2, and MACRO on 
$\nu_\mu + \bar \nu_\mu$ events through the earth should be analyzed for 
such an effect.  For a zenith angle near zero, the $\Delta m^2 \sim 10^{-2}$ 
eV$^2$ oscillation scenario should have $R \sim 1$, whereas the $\Delta m^2 
\sim 0.25$ eV$^2$ oscillation scenario (with anomalous interaction) would 
have $R = P_0 \sim 0.66$.  Furthermore, if $\nu$ and $\bar \nu$ can be 
distinguished (as proposed in the HANUL experiment), then to the extent 
that $CP$ is conserved, matter effects can be isolated.

\section*{Acknowledgments} I thank Gustavo Branco and all the other organizers 
of this stimulating School and Workshop for their hospitality.  This work was 
supported in part by the U.~S.~Department of Energy under Grant 
No.~DE-FG03-94ER40837.

\end{document}